\documentclass[letterpaper]{article}
\usepackage{spconf,amsmath,amssymb,graphicx}
\usepackage[english]{babel}
\usepackage{algpseudocode}
\usepackage{algorithm}
\usepackage{tikz}
\usetikzlibrary{decorations.pathreplacing}

\newcommand{\mypar}[1]{{\bf #1.}}

\title{Hierarchization for the Sparse Grid Combination Technique
}
\name{Philipp Hupp} 
\address{Department of Computer Science\\ ETH Z\"urich\\Z\"urich, Switzerland}

\begin{document}
	%
	\maketitle

	\begin{abstract}
		The sparse grid combination technique provides a framework to solve high dimensional numerical problems with standard solvers. 
		Hierarchization is preprocessing step facilitating the communication needed for the combination technique. 
		The derived hierarchization algorithm outperforms the baseline by up to 30x and achieves close to 5\% of peak performance. It also shows stable performance for the tested data sets of up to 1 GB.

	\end{abstract}
	
	\section{Introduction}\label{sec:intro}

		\mypar{Motivation}
			 The standard way to discretize Euclidean spaces for scientific computations, for example to solve partial differential equations, are regular grids. However, in $d$ dimensions classical full grid discretizations need $n^d$ grid points, where $n$ represents the number of grid points per axis. This exponential dependency of the amount of data on the dimension is called the curse of dimensionality. Because of this phenomena, full grid discretizations cannot be employed for high dimensions, namely \mbox{$ d>4$}. 


			Sparse grids \cite{smolyak63quadrature, zenger91sparse, bungartz04sparseGrids} are a discretization method reducing the curse of dimensionality from  $n^d$ to  $\mathcal{O}\left(n \cdot \left(\log n\right)^{d-1}\right) $. Hence problems in higher dimensions are feasible with this technique.  As a drawback sparse grids introduce (more complicated) data dependencies to non-neighboring grid points.

			The sparse grid combination technique \cite{griebel92CombiTechnique} provides a workaround for this issue: The sparse grid is decomposed into several anisotropic full grids called \emph{combination grids}. The problem is then solved on these combination grids with standard solvers in parallel. This additional, very coarse level of parallelism also makes the combination technique interesting for exascale computing.
			The sparse grid solution can then be approximated by calculating a weighted sum of the solutions on the combination grids. Therefore the combination grids need to exchange their information in a \emph{communication phase}.
			
			The \emph{hierarchization} algorithm optimized in this paper is a preprocessing step facilitating the communication phase. 

			This preprocessing step is in particular important for the iterated combination technique: After assembling the sparse grid solution it is projected back onto the combination grids. Then, another round of solving on the combination grids, gathering the sparse grid solution and scattering the sparse grid solution back onto the combination grids starts. Hence the communication phase is integrated into an iterative setup. This stresses the need for an efficient implementation of the communication phase and the preprocessing step hierarchization as a speedup in the overall algorithm can only be expected if the overhead created by the communication phase is less than the savings in the compute phase.

\noindent		\mypar{Related Work}
			Sparse grids \cite{smolyak63quadrature, zenger91sparse, bungartz04sparseGrids} and the sparse grid combination technique \cite{griebel92CombiTechnique} have been used successfully to solve high dimensional problems. Also, the convergence of the sparse grid solution can be proven. In contrast, the convergence of the combination grid solution approximating the spare grid solution is much harder to prove. Even when the coefficients used to combine the combination grids are choose optimally \cite{hegland07optiCom} only few results are known.
			No convergence results are known for the iterated combination technique and few experiments have been conducted \cite{griebel96numericalTurbulence}. However, these experiments show promising results. 

			While several software packages to hierarchize spare grids have been developed, few of these can handle the regular grids of the combination technique \cite{pflueger10spatially,buse12non-static}. We use \emph{SGpp} \cite{pflueger10spatially}, the current standard for sparse grids, as baseline against which we benchmark our code. \emph{SGpp} solves a more general problem as it can deal with spatially adaptive sparse grids. We restrict ourselves to the anisotropic grids of the sparse grid combination technique. This makes navigating on the data layout much easier and should result in significant performance gains. Besides \emph{SGpp}, \cite{buse12non-static} provides a reportedly fast implementation of the hierarchization algorithm which can handle combination grids. Time did not allow to benchmark against this code.

			All experiments were preformed using the roofline tool~\cite{ofenbeck13applyingRoofline} for the measurements.

		\section{Sparse Grids and the (Iterated) Combination Technique}

		This section first describes how sparse grids lessen the curse of dimensionality. Then the sparse grid combination technique and the iterated combination technique are discussed. We conclude by describing the role of the hierarchization algorithm as preprocessing step for the communication phase.

		\noindent	\mypar{Sparse Grids}
			\emph{Sparse Grids} reduce the curse of dimensionality by combining a tensor product approach with a base change from the usual full grid basis (in case of piecewise linear functions also called nodal basis)  to the hierarchical basis. This representation allows the sparse grid to select the most important variables a priori.

			The hierarchical basis introduces (more complicated) data dependencies to non-neighboring grid points. Therefore, numerical problems are harder to solve as the algorithms have to account for the altered data dependencies. As a result not a lot of algorithms have been implemented to work directly in the hierarchical basis of sparse grids.


		\noindent \mypar{The Combination Technique}
			The sparse grid combination technique \cite{griebel92CombiTechnique} provides a workaround for this issue: The sparse grid is decomposed into several, $\mathcal{O}\left(d \cdot l^{d-1}\right) $, an\-isotropic, i.e. refined differently in different dimensions, full grids called \emph{combination grids}. Hereby $l$ is the maximum refinement level which is related to the number of grid points per axis by roughly $l \approx \log n\enspace$. A combination grid is completely described by its level vector $\ell \in \mathbb{N}^d$. For $1 \leq i \leq d$, $\ell_i$ describes how often dimension $i$ has been refined. We work with the convention that a grid that has refinement level 1 consist of one single grid point.

\begin{figure}[htbp]
				\includegraphics[width=\columnwidth]{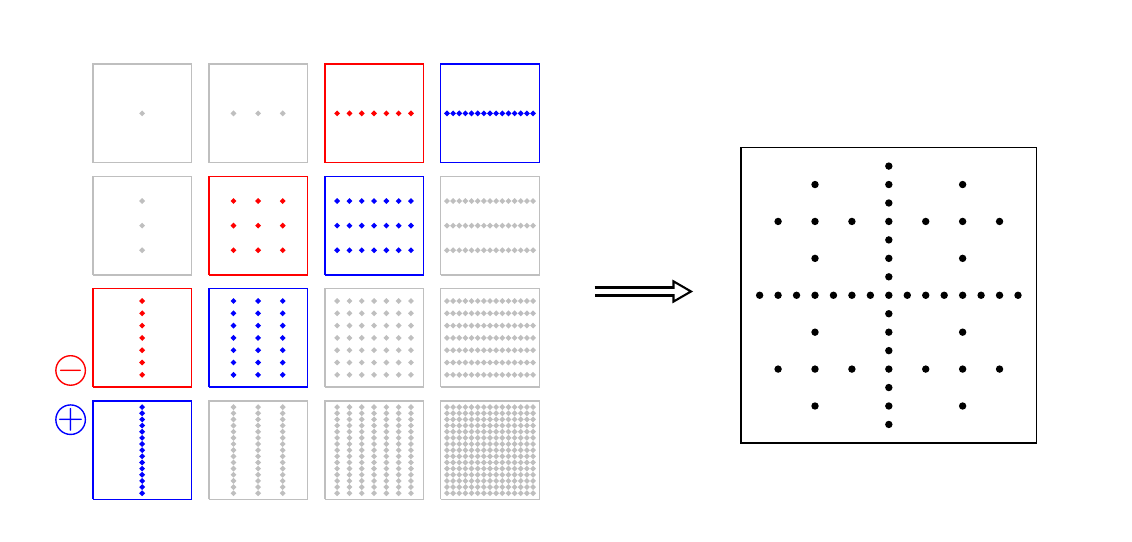}
				\caption{In two dimensions: Left: Combination grids of different refinement level. For the combination technique the solutions on the blue combination grids are weighted with $+1$, the solutions on the red ones are weighted with $-1$. Right: The sparse grid resulting from the combination technique.}
				\label{fig:CGsAndSG}
			\end{figure}

			The problem of interest is solved on the combination grids with standard solvers.
			The sparse grid solution can then be approximated by calculating a weighted sum of the solutions on the combination grids as depicted in Figure \ref{fig:CGsAndSG}.
			Solving the problem of interest on the combination grids allows to use usual regular full grid algorithms as black box.  Hence it is not necessary to design algorithms especially for sparse grids. Furthermore an additional, very coarse level of parallelism is introduced as the solutions on the combination grids can be computed completely independently.

			The drawback that accompanies the parallelism during this \emph{compute phase} is the need to communicate between the combination grids to assemble the sparse grid solution. We call this the \emph{communication phase} of the sparse grid combination technique. As the focus of sparse grids are high dimensions and data intense settings, the number of combination grids, the number of grid points of one combination grid or both may be huge. Hence the communication phase may pose a bottleneck for the combination technique.

\begin{figure}[htbp]\centering
				\includegraphics[width= \columnwidth]{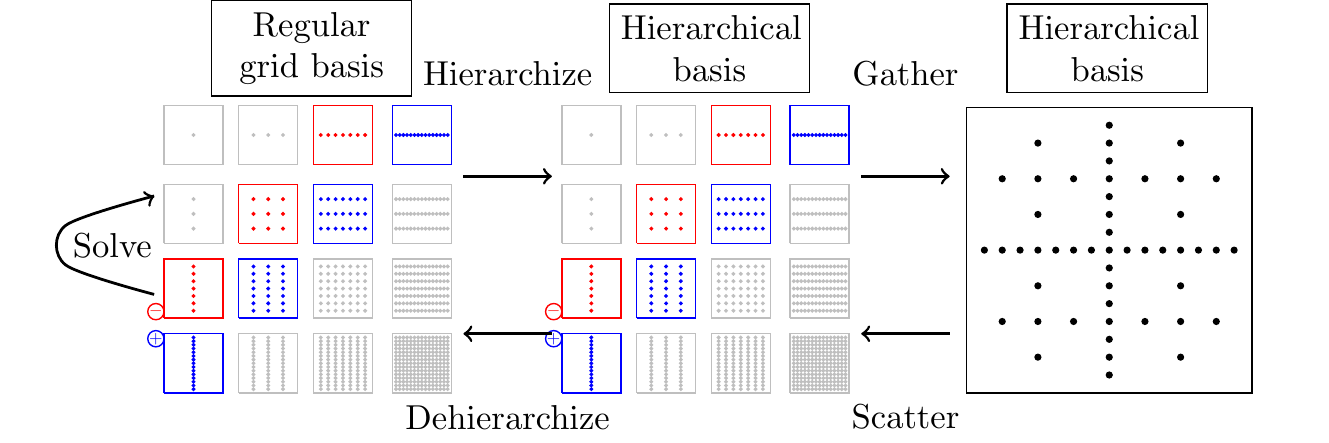}
				\caption{The iterated combination technique: Performing a base change from the regular grid basis to the hierarchical basis facilitates the following communication step.}
				\label{fig:iterCombiHier}
			\end{figure}

		\noindent \mypar{The Iterated Combination Technique}
			Besides accumulating the global sparse grid solution when the solution on all combination grids has been computed, we aim for an iterated combination technique \cite{griebel96numericalTurbulence} (see Fig.~\ref{fig:iterCombiHier}):
			On each of the combination grids an iterative solver is used to compute $t$ time steps. Then all combination girds are hierarchized in parallel before the sparse grid solution is assembled in a gather step. Afterwards, this sparse grid solution is projected onto and distributed to every combination grid in a scatter step. Then the combination grids are dehierarchized, transforming the function values from the hierarchical back to the regular grid basis.  This process is repeated and continues with the next $t$ time steps of the iterative solver on the combination grids. 

			The goal is to reduce the number of iterations needed by the iterative solver on the combination grids by exchanging information between combination grids. Ideally, as the combination grids are refined differently, numerical errors on different grids should cancel and allow for a faster convergence of the solution. 

			\noindent \mypar{Hierarchization as preprocessing}
			To facilitate the communication as much as possible we hierarchize before communicating.
			Consider the combination grids depicted in Fig. \ref{fig:iterCombiHier} and think of each grid as representing a function on the global domain. As the combination grids are differently refined, combining the solution of two of them requires an interpolation and sampling process to get the function values of one solution at all grid points of the other combination grid. If the coefficients are given in the hierarchical basis, all grid points that do no exist in a combination grid have a coefficient, also called hierarchical surplus, of 0. Hence interpolation is no longer necessary.

	\section{The Hierarchization Algorithm}\label{sec:yourmethod}

		In this section the hierarchization algorithm is explained and the number of flops it has to execute are derived. Afterwards the different implemented versions of the hierarchization algorithm are described. All versions have been verified with the standard software \emph{SGpp}.

		\noindent \mypar{The Hierarchization Algorithm}
			On a abstract level the hierarchization algorithm is described in Algorithm \ref{alg:hierarch}. The outer loop iterates over the $d$ dimensions. The inner three loops update the whole data set once. Hereby, the second loop splits the data set into 1-dimensional poles in the current working dimension. The inner two loops perform the updates on these 1-dimensional poles in a \emph{daxpy} like fashion. 
			\begin{algorithm}[htbp]
				\begin{algorithmic}
				\For {$d \gets  1,\; \dots\; ,\;\text{dim}$} 
					\For{all 1-dim poles in direction $d$}
						\For{$\ell \gets \ell_d,\;\dots\; ,\;2$ }
							\For{ all $x_i$ on level l}  
								\State $x[i] = x[i] - 0.5 * \text{leftPredecessor}(i,\; d)$
								\State $x[i] = x[i] - 0.5 * \text{rightPredecessor}(i,\; d)$
							\EndFor
						\EndFor		
				\EndFor
				\EndFor
				\end{algorithmic}
				\caption{Hierarchization of a $dim$ dimensional Combination Grid of level vector  $(l_1,l_2,\dots,l_d)$.}
				\label{alg:hierarch}
			\end{algorithm}

\begin{figure}[htbp]
				 \renewcommand{\tabcolsep}{0.12cm}

				\begin{tabular}{ccc}

				\includegraphics[width=.4\columnwidth]{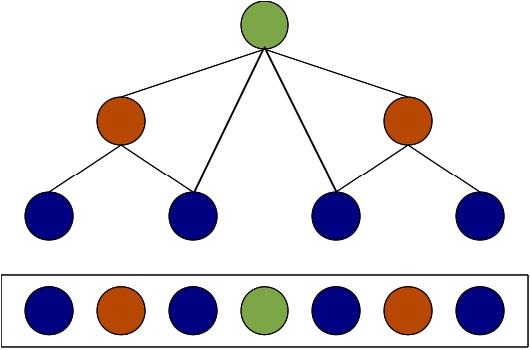} 
				& 
				\begin{tabular}{l}
				\vspace{-15ex}\\
				BFS Layout\\
				\begin{tikzpicture}[scale=1]
					\foreach \x in {3,...,6}{
						\draw[fill, blue] (\x/4,0) circle (0.1);
					}
						\draw[fill, red] (1/4,0) circle (0.1);
						\draw[fill, red] (2/4,0) circle (0.1);
						\draw[fill, green] (0/4,0) circle (0.1);
					\draw[thick] (-0.15,-0.15) rectangle (1.65,0.15);
				\end{tikzpicture}
				\vspace{1ex}\\
				
				Reverse \emph{BFS} \\Layout \\
				\begin{tikzpicture}[scale=1]
					\foreach \x in {0,...,3}{
						\draw[fill, blue] (\x/4,0) circle (0.1);
					}
						\draw[fill, red] (4/4,0) circle (0.1);
						\draw[fill, red] (5/4,0) circle (0.1);
						\draw[fill, green] (6/4,0) circle (0.1);

					\draw[thick] (-0.15,-0.15) rectangle (1.65,0.15);
				\end{tikzpicture}
			\end{tabular}
			& 	\begin{tikzpicture}[scale=1.2]
					\foreach \y in {1,...,7}{
						\foreach \x in {0,...,3}{
							\draw[fill, blue] (\y/4,\x/4) circle (0.1);
						}
							\draw[fill, red] (\y/4,4/4) circle (0.1);
							\draw[fill, red] (\y/4,5/4) circle (0.1);
							\draw[fill, green] (\y/4,6/4) circle (0.1);
					}
					\draw[black,thick] (1/8,-1/8) rectangle (1.12,1/8);
					\draw[black,thick,yshift = 1cm] (1/8,-1/8) rectangle (1.12,1/8);
					\draw[black,thick,yshift = 1.5cm] (1/8,-1/8) rectangle (1.12,1/8);

					\draw[orange,very thick] (1/8,-1/8) rectangle (3/8,13/8);

				\end{tikzpicture}
			\end{tabular}
				\caption{\textbf{Left:} A 1-dimensional grid or pole in the regular grid layout. Hierarchical predecessors are above their children in this binary tree like structure. \textbf{Middle:} The same grid points (1-dimensional poles) in \emph{BFS} and reverse \emph{BFS} layout. \textbf{Right:} Working orthogonal the poles (orange) enables vectorization in $x_1-$direction (vector registers black).}
				\label{fig:hierDehier}

			\end{figure}
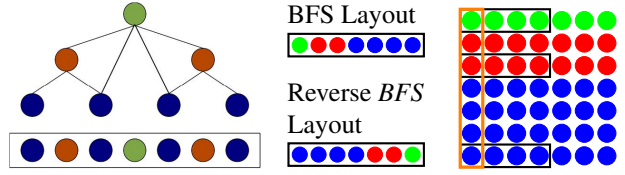

			It is left to describe the hierarchical predecessors for the 1-dimensional poles or grids. Think of a 1-dimensional grid as a hierarchical, binary-tree like structure as depicted in the left part of Fig.~\ref{fig:hierDehier}. The left (right) hierarchical predecessor of a node $v$ is above $v$ in this tree-like structure and the closest vertex on the left (right) of $v$ in the 1-dimensional projection of the structure. Every vertex besides the root has at least one hierarchical predecessor. The second hierarchical predecessor does not exist for the outermost grid points of each refinement level.

%

			\noindent \mypar{Flop Count}
			\label{sec:OpCount}
			The flop $F$ count has been deduced from Alg.~\ref{alg:hierarch} and the derivations have been verified by instructing the code.
			To hierarchize  a $d$ dimensional combination grid of level vector $(l_1,l_2,\dots,l_d)$ 
			\begin{equation}
			F(d,\ell) \!= \!2\cdot \! \sum_{i=1}^d\! \!\left(\!\left(2^{l_i}-2\cdot l_i-2 \right) \!\cdot \!\!\!\!\!\!\prod_{j=1, \; j \neq i}^d \!\left(2^{l_j}-1\right)\! \right) \, .
			\label{eq:flops}
			\end{equation}
			flops are needed. These split equally into additions and multiplications.


			The flop count of the hierarchization can be reduced. Whenever both hierarchical predecessors exist, their values are first added, then multiplied by $-0.5$ and then added to the value of $x[i]$. This saves 1 multiplication operation but no addition. This reduces the number
			 of multiplications to
			\begin{equation*}
			 M(d,\ell)=  \sum_{i=1}^d \left(\left(2^{l_i}-2 \right) \cdot \prod_{j=1, \; j \neq i}^d \left(2^{l_j}-1\right) \right) 
			\end{equation*}
			while the number of additions stays at \mbox{$ A(d,\ell) = F(d,\ell)/2 $}\enspace.
			As we then do roughly twice as many additions as multiplications the reachable peak performance should be $75\%$ of the theoretical peak performance.
			
			\noindent \mypar{Chosen results about reducing the flop count}
			For the \emph{Ind} layout, where both hierarchical predecessors are equally easy to compute, reducing the flop count did not change the number of cycles needed.

		\noindent \mypar{Baseline using \emph{level-index} vector}
			As baseline the \emph{Func} algorithm navigating on the combination grids using a \emph{level-index} vector as in the baseline \emph{SGpp} was implemented. The grid data is stored in standard row major oder. For each dimension the \emph{level-index} vector describes on which refinement level the current grid point is (level) and which position the grid point takes on that level (index). \emph{SGpp} has a large memory footprint since it provides memory to adaptively refine the grid. Therefore we could only run it for small problem instances. The implementation \emph{Func} provides a baseline for all tested input sizes.

		\noindent \mypar{Indirect navigation on the data layout}
			As the combination grids are very regular the level index-vector is not necessary to navigate efficiently on the data layout. The \emph{Ind} algorithm navigates indirectly on the data layout in the sense that it does not  need the \emph{level-index} vector. The positions of the hierarchical predecessors and the next grid point can be computed on the fly by using offsets and strides.
 
		\noindent \mypar{BFS layouts}
			As we access the 1-dimensional poles  bottom up level by level (see Alg.~\ref{alg:hierarch}), the \emph{BFS} and \emph{Reverse-BFS} layouts (Fig.~\ref{fig:hierDehier}) organize the data according to the levels. The ordering of the grid points  can be seen as a \emph{BFS} (breadth first search) traversal of the binary-tree like structure shown in Fig.~\ref{fig:hierDehier}.

		\noindent \mypar{Unrolling and Vectorization}
			When working in the direction of the data layout vectorization is difficult. However, whenever the poles are aligned orthogonal to the fastest changing index ($x_1$ for row major order) any regular data layout is suitable for vectorization (and parallelization) (see Fig.~\ref{fig:hierDehier}). All poles can be handled independently and the data of neighboring poles are contiguous in memory. For the experiments the code has first been unrolled by a factor of 4 resulting in the \emph{BFS-Unrolled} code. Afterwards manual vectorization using \emph{AVX} was employed resulting in code \emph{BFS-Vectorized}. Only the algorithms working in the \emph{BFS} layout have been vectorized, as they achieve the highest performance for large input sizes. For moderate input sizes vectorized versions of the \emph{Ind} code should be beneficial. 

			To work only with aligned loads and stores one grid point needed to be padded to every pole pointing in the first dimension.

		\noindent \mypar{Over-vectorization, pre-branching and reducing the op-count}
			All poles of one working direction can be handled in parallel. If the working direction is at least 2, we unrolled (and vectorized) the innermost loop such that $2^{l_1}-1$ poles are handled instead of a single one in the innermost loop (\emph{BFS-OverVectorized}).
			This unrolling also allows to decide the branch if two or only one hierarchical predecessor are present for  $2^{l_1}-1$ poles at once (\emph{BFS-OverVectorized-PreBranched}). 
			The optimization tried last was to reduce the operation count as described previously in this section (\emph{BFS-OverVectorized-PreBranched-ReducedOp}).

	\section{Experimental Results}\label{sec:exp}

	The effects of navigation overhead, different data layouts and vectorization onto the performance of the hierarchization algorithm are examined in this section.

	\noindent \mypar{Experimental setup}
	All experiments were performed on a SandyBridge Corei7-2620M  running at 2.7GHz. TurboBoost was disabled.  The L1-cache is 64KB, the L2-cache  256 KB, the L3-cache 4MB and the main memory 4 GB large. The largest data set examined was 1 GB. We work with 1 GB of data when the levelsum $|\ell|_1 = 27$. If the levelsum decreases by one, the size of the data set halves. All experiments were performed for double precision data and for vectorization the 4-way \emph{AVX} registers were used.

	As compiler \emph{icc} version 13.0.0 was employed and the flags \emph{-std=c99 -xHost -O3 -no-simd -no-vec} were used.

	For the roofline plots \cite{williams09roofline} the memory bandwidth was taken from the stream benchmark~\cite{calpin2007streamRep}. The compute bound is always depicted as scalar peak performance although it does not apply to the vectorized versions.

\begin{figure}[htbp]\centering
				\includegraphics[width=\columnwidth]{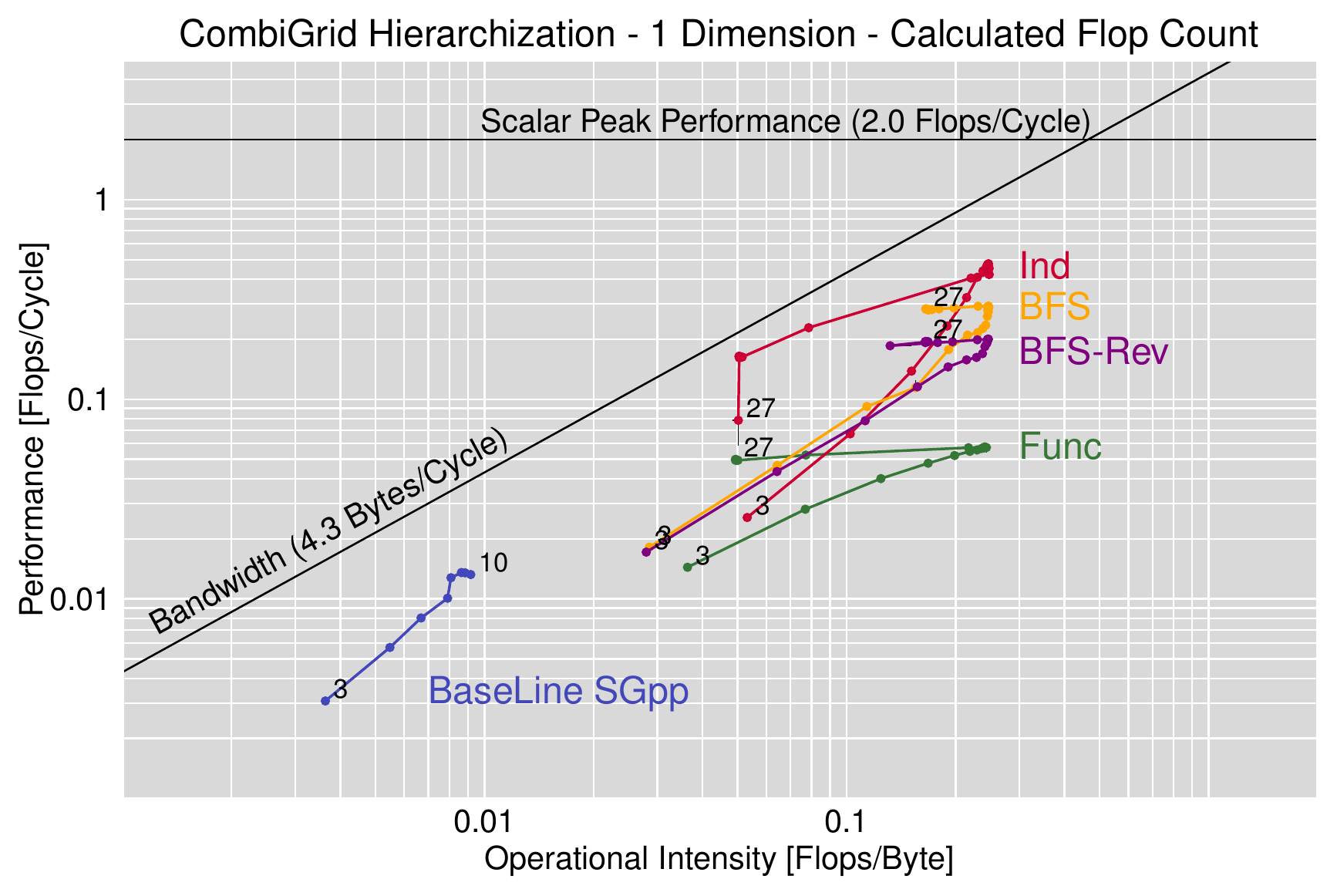}
				\caption{Hierarchizing a 1-dimensional grid. Performance for calculated flop count.}
				\label{fig:1DLayouts}
			\end{figure}

	\noindent \mypar{The data layout matters for large poles}
	The first experiment (Fig.~\ref{fig:1DLayouts}) shows the performance of different data layouts for a one dimensional grid. To calculate the performance the theoretically deduced flop count from Eq.~\ref{eq:flops} was used.
	For $l = 27$ the data sets are about 1 GB large. The \emph{Ind} layout, navigating on a usual grid layout without the \emph{level-index} vector, achieves best performance for moderate input size (up to about 100 MB). For larger input the performance drops and the \emph{BFS} layouts achieve better performance. The performance of the \emph{BFS} algorithms stays constant as the size of the data set increases further. The \emph{BFS} algorithm is about 50\% faster than the \emph{BFS-Rev} algorithm. Hence no further experiments were conducted with the latter. It can be observed that all implementations beat the baseline \emph{SGpp} and the implementation \emph{Func} is beaten by all implementations besides \emph{SGpp}.

\begin{figure}[htbp]\centering
				\includegraphics[width=\columnwidth]{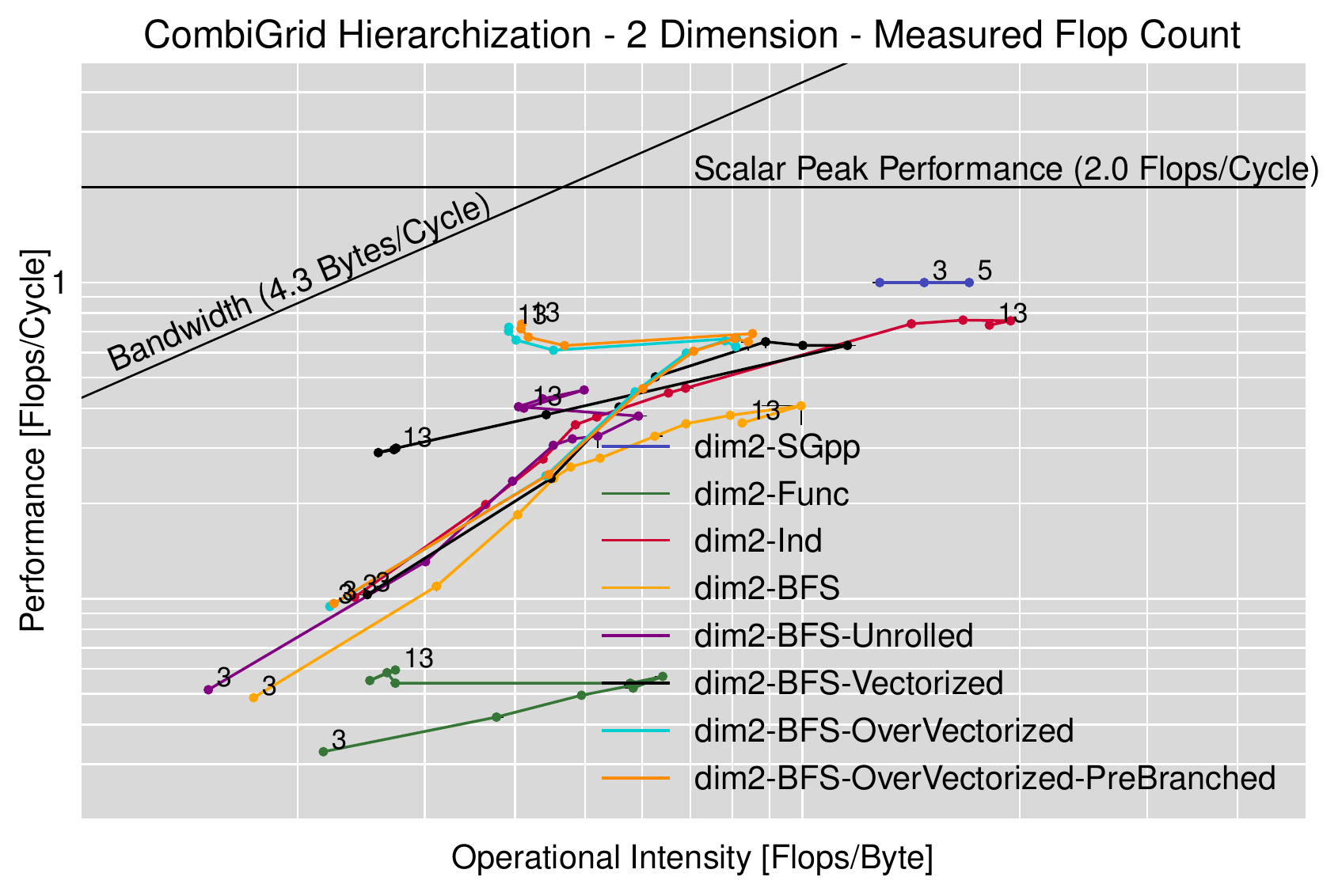}
				\caption{Measured performance for two dimensional grids.}
				\label{fig:2DMeasured}
			\end{figure}

\begin{figure}[htbp]\centering
				\includegraphics[width=\columnwidth]{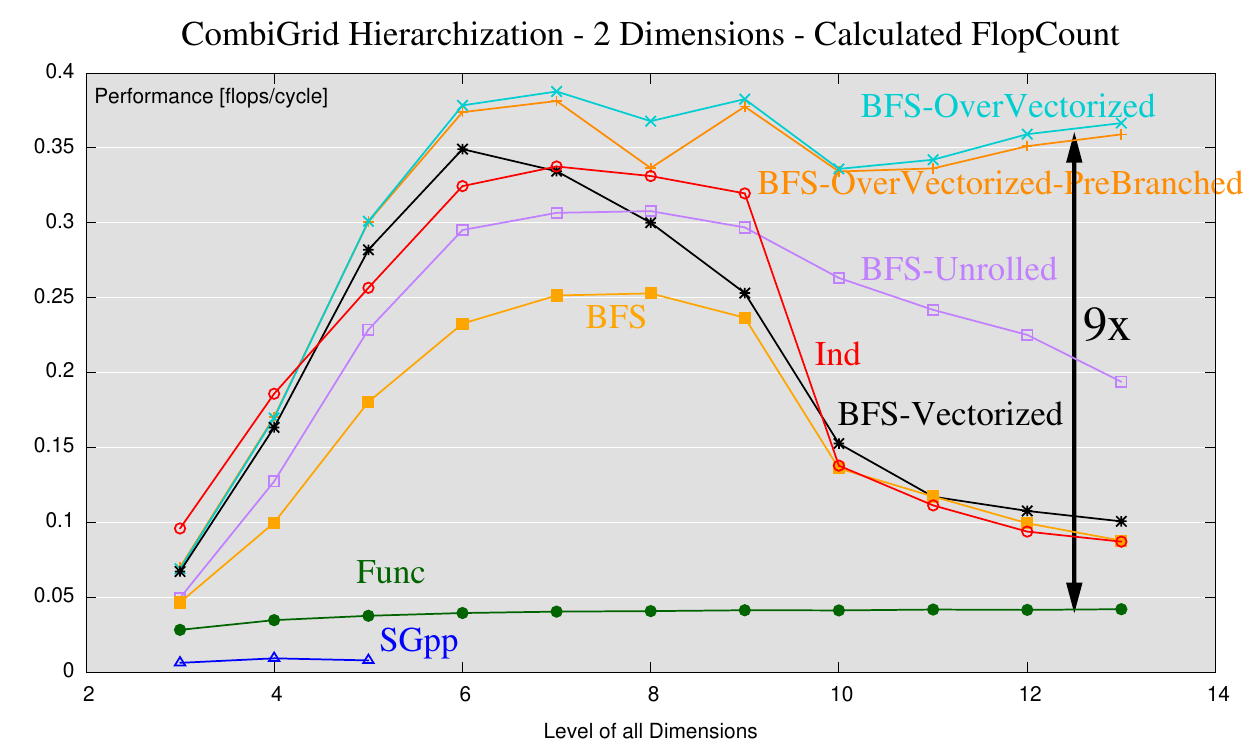}
				\caption{Calculated performance for two dimensional grids.}
				\label{fig:2DCalculated}
			\end{figure}

	\noindent \mypar{Measuring performance may point the wrong way}
	The second experiment hierarchizes two dimensional grids of different size and compares measured performance (Fig. \ref{fig:2DMeasured}) with performance derived using the calculated flop count from  Eq.~\ref{eq:flops} (Fig.~ \ref{fig:2DCalculated}). 

	Although \emph{SGpp} seems to achieve the highest performance for the measured case it actually performs worst for the calculated case that directly mirrors wall clock time. Navigating on the data structure can be done by integer operations. Hence we want to disregard the overhead of navigating on the data structure. Non-optimal code may use floating point operations for this navigation and hence pretend better performance. However, this performance gain would not yield any runtime improvements. 

	Also the performance of the \emph{Ind}-algorithm seems to be to high when measured. For the \emph{Ind}-code this may be due to branch mispredictions which may lead to flops that are not relevant for the results. 
	Therefore the upcoming experiments will use Eq.~\ref{eq:flops} to derive performance from the measured number of cycles taken for the computation.

	One can also observe that roofline plots are difficult to read due to the log-log-scaling if the performance results  are of the same order of magnitude.

\begin{figure}[htbp]\centering
				\includegraphics[width=\columnwidth]{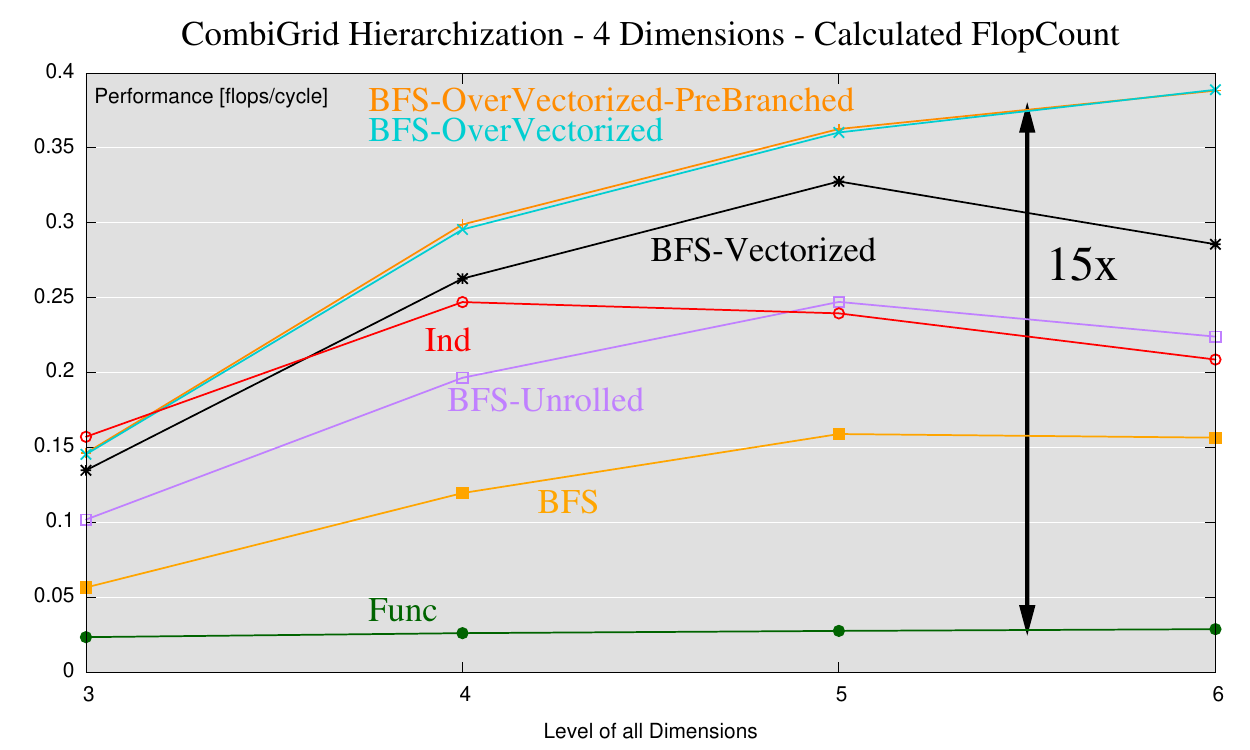}
				\caption{Hierarchizing a 4 dimensional grid.}
				\label{fig:4D}
			\end{figure}

\begin{figure}[htbp]\centering
				\includegraphics[width=\columnwidth]{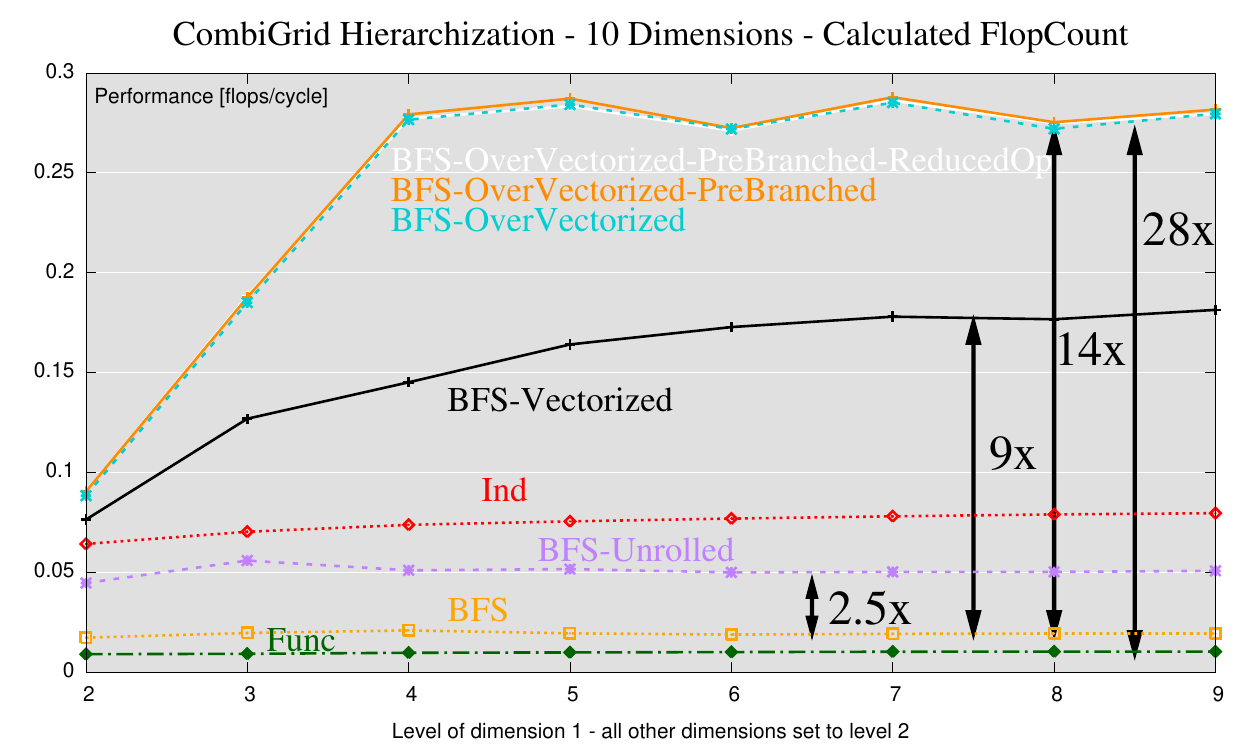}
				\caption{Hierarchizing a 10 dimensional anisotropic grid. The number of points of the first dimension are increased while all other dimensions are fixed to 3 grid points.}
				\label{fig:10D}
			\end{figure}

	\noindent \mypar{Vectorizing and Over-Vectorizing}
	When the data is stored in row major order, all dimensions except the first one are nicely laid out for vectorization. 
	Figure~\ref{fig:2DCalculated} (2 dimensional), Fig.~\ref{fig:4D} (4 dimensional) and Fig.~\ref{fig:10D} (10 dimensional) show that unrolling (by a factor of 4) and then vectorizing the code yield significant performance gains. 
	 \emph{BFS-OverVectorization} further increases the performance.  However, deciding the hierarchical predecessor branch for several poles at once (\emph{BFS-OverVectorized-PreBranched}) did not yield any further performance gains.

	\noindent \mypar{Reducing the flop count}
			When the flop count is reduced as described in Sect.~\ref{sec:OpCount} the critical path remains three flops long but  both hierarchical predecessors take then part in this critical path. Before reducing the flop count, the second hierarchical predecessor took only part in two floating point operations. For the \emph{BFS} layouts, calculating the hierarchical predecessors requires branching to navigate in the tree like structure of Fig.~\ref{fig:hierDehier}. In particular, one predecessor is directly one level above the current node while the other may require to traverse the tree up to the root. When the flop count is reduced this predecessor has to take part in three instead of two flops and hence no runtime gain was expected (and measured -- the experiment is not presented here). 
%
%

			For the \emph{BFS-OverVectorization} code the computation of the hard hierarchical predecessor is done for $2^{l_1}-1$ poles at once but even in this case (\emph{BFS-OverVectorized-PreBranched-ReducedOp}) no runtime improvements have been observed (see Fig.~\ref{fig:10D}).

\begin{figure}[htbp]\centering
				\includegraphics[width=\columnwidth]{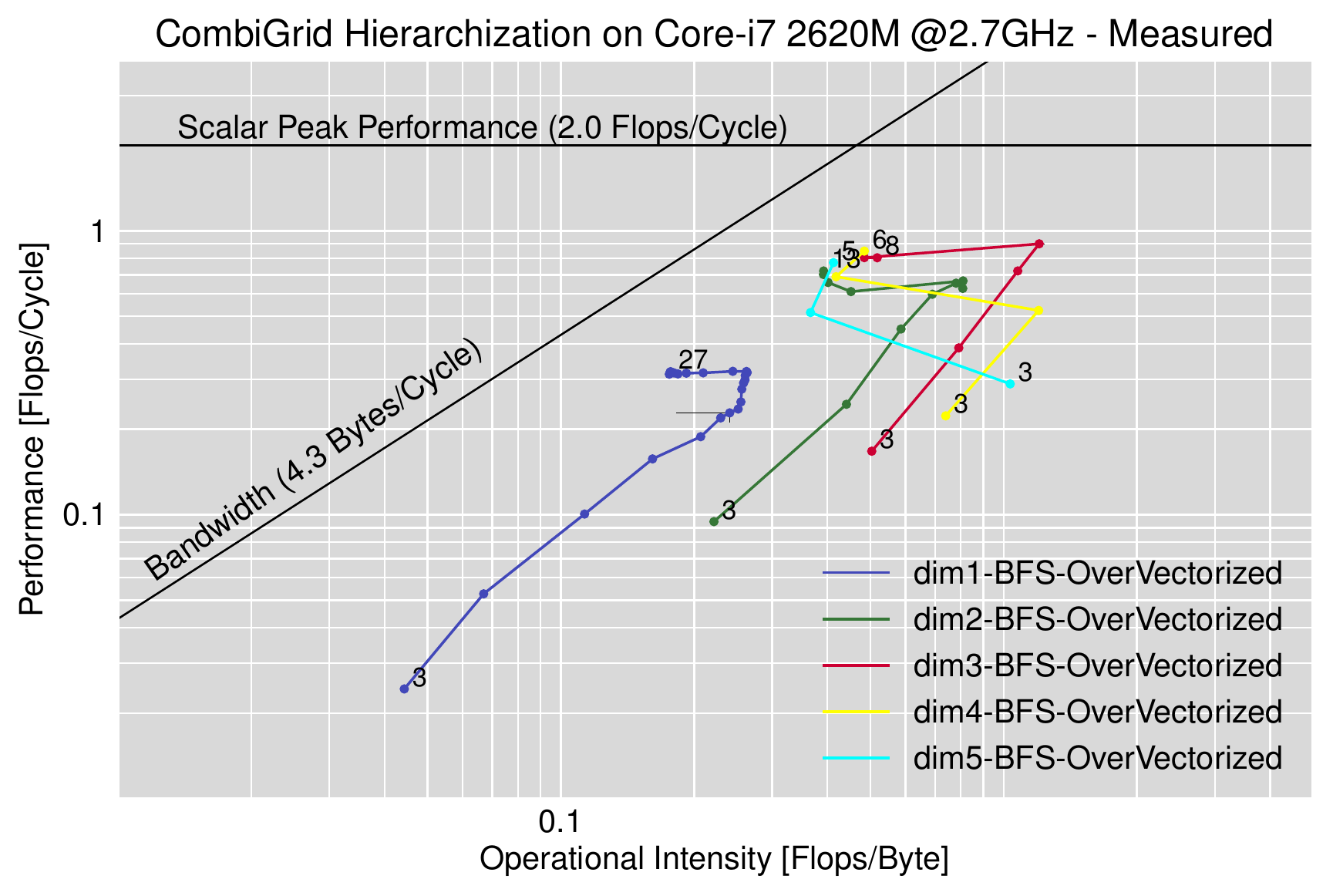}
				\caption{Measured performance of \emph{BFS-OverVectorization} in different dimensions.}
				\label{fig:BFSOver}
			\end{figure}

	\noindent \mypar{Performance of the \emph{BFS}-OverVectorized Layout over different dimensions}
		The measured, not calculated, performance of the best code (\emph{BFS-OverVectorization}) running in different dimensions is shown in Fig.~\ref{fig:BFSOver}. For $2 \leq d \leq 5$ the maximum grid sizes are roughly the same and the grids are between 125 MB and 500 MB large. The achieved performance is very similar for these dimensions and only lower for the 1-dimensional case. Also the achieved operational intensity is very similar for large grids when the dimension is between 2 and 5. Hence this code should perform well, even if the first dimension has not been refined a lot.

	\section{Summary of he Experimental Results}

	The best code implemented, \emph{BFS-OverVectorized}, achieves up to $0.4$ flops per cycle or 5\% of the peak floating point performance using \emph{AVX} and double precision. This is a speedup between 10x and 30x against the baseline implementation \emph{Func}. This baseline always outperforms the standard sparse grid software \emph{SGpp} by another factor between 2x  (Fig.~\ref{fig:10D}) and 10x (Fig.~\ref{fig:1DLayouts}).

	It is also worthwhile noting that the performance of \emph{BFS-OverVectorized} and in particular \emph{BFS} stays constant as the input size increases to up to 1 GB of data.

	\section{Further Ideas}
		This section gathers further ideas about the project.
		
		As \emph{SGpp} has been designed to account for spatially adaptive sparse grids it is necessary to benchmark the derived code against \cite{buse12non-static} which has been optimized for performance.

		Vectorizing the \emph{Ind} layout should yield an algorithm outperforming the vectorized version of the \emph{BFS} algorithm for moderate input sizes as.

		Neither different compiler flags nor different compilers were tested. Also, compiler vectorization was always disabled. This did not matter as the data access pattern was to complicated for compiler vectorization before over-vectorization was implemented. With this approach, however, compiler vectorization could be possible and this should be examined. 

		Rotating the data would destroy the layout used for vectorization and over-vectorization and hence does no longer seem like an approach worthwhile trying. Also copying the data into contiguous is no longer necessary as neighboring poles are already in contiguous memory.



	\bibliographystyle{../../../TeX/IEEEbib}
	\bibliography{../../bibPhilipp}

\end{document}